\pdfoutput=1

\documentclass[11pt]{article}

\usepackage{EACL2023}

\usepackage{times}
\usepackage{latexsym}

\usepackage[T1]{fontenc}

\usepackage[utf8]{inputenc}

\usepackage{microtype}

\usepackage{inconsolata}

\usepackage{graphicx}
\usepackage{booktabs}
\usepackage{multirow}

\usepackage{listings}
\usepackage{makecell}

\title{Context Tuning for Retrieval Augmented Generation}

\author{Raviteja Anantha, Tharun Bethi, Danil Vodianik, Srinivas Chappidi \\
Apple}

\begin{document}
\maketitle
\begin{abstract}
Large language models (LLMs) have the remarkable ability to solve new tasks with just a few examples, but they need access to the right tools. Retrieval Augmented Generation (RAG) addresses this problem by retrieving a list of relevant tools for a given task. However, RAG's tool retrieval step requires all the required information to be explicitly present in the query. This is a limitation, as semantic search, the widely adopted tool retrieval method, can fail when the query is incomplete or lacks context. To address this limitation, we propose Context Tuning for RAG, which employs a smart context retrieval system to fetch relevant information that improves both tool retrieval and plan generation. Our lightweight context retrieval model uses numerical, categorical, and habitual usage signals to retrieve and rank context items. Our empirical results demonstrate that context tuning significantly enhances semantic search, achieving a 3.5-fold and 1.5-fold improvement in Recall@K for context retrieval and tool retrieval tasks respectively, and resulting in an 11.6\% increase in LLM-based planner accuracy. Additionally, we show that our proposed lightweight model using Reciprocal Rank Fusion (RRF) with LambdaMART outperforms GPT-4 based retrieval. Moreover, we observe context augmentation at plan generation, even after tool retrieval, reduces hallucination.

\end{abstract}

\section{Introduction}
Large language models (LLMs) excel in a variety of tasks ranging from response generation and logical reasoning to program synthesis. One of the important active areas of LLM research is to utilize them as planning agents~\cite{llm-planner:22}. Planning is an essential functionality for processing complex natural language instructions. A planner should possess the ability to select the appropriate tools to complete each sub-task. While LLMs exhibit exceptional generation capabilities, they have inherent limitations, such as lacking up-to-date information and exhibiting a tendency to hallucinate tools. By providing LLMs with a relevant set of tools based on the given task~\cite{toolformer:2023, chameleon:2023}, one can alleviate the issue of outdated information. The set of methods to augment LLM input with retrieved information, such as relevant tools, is referred to as Retrieval Augmented Generation (RAG)~\cite{realm:20, rag-nlp:20}. RAG consists of three primary components: Tool Retrieval, Plan Generation, and Execution.\footnote{Typically, the query along with retrieved tools undergo dynamic prompt construction before presented to an LLM. This process is called Query Decoration/Transformation. We omit that in this work for the sake of simplicity.} In this study, we focus on enhancing tool retrieval, with the goal of achieving subsequent improvements in plan generation.



Existing RAG methodologies rely heavily on semantic search for tool retrieval, but this approach has limitations, especially when queries lack specificity or context. To this end, we present Context Tuning, a component in RAG that precedes tool retrieval, to provide contextual understanding and context seeking abilities to improve tool retrieval and plan generation. Our contribution can be summarized as follows:
\begin{enumerate}
  \item We empirically show that traditional RAG is inadequate for implicit/context-seeking queries and present context tuning as a viable solution;
  \item We provide a systematic comparison of various context retrieval methods applied on both lightweight models and LLMs;
  \item We share empirically the insight that Chain of Thought (CoT) augmentation improves context retrieval when no fine-tuning is applied, whereas fine-tuning the retrieval model removes the need for CoT augmentation;
  \item We propose a lightweight model using Reciprocal Rank Fusion (RRF)~\cite{rrf:09} with LambdaMART~\cite{lambdamart:2010}, which  outperforms GPT-4~\cite{gpt-4:23} system, and finally;
  \item We show that context augmentation at plan generation reduces hallucinations.
\end{enumerate}

\section{Related Work}
Using retrieval to incorporate tools into plan generation with LLMs has emerged as a burgeoning area of research, with ongoing investigations aimed at enhancing both the retrieval component and the LLMs themselves. Our work falls within the former category, placing a particular emphasis on refining retrieval methodologies to enhance contextual understanding of implicit and ambiguous queries that demand context-seeking capabilities.

The integration of tools into generation has been demonstrated to enhance the capabilities of LLM-based planners in recent studies~\cite{toolformer:2023, chameleon:2023}. However, these works primarily focus on well-defined or unambiguous queries, where retrieving supplementary information to augment the query is not strictly required. For question answering (QA) tasks, incorporating any off-the-shelf document retriever has been shown to improve LLM generation, with the addition of re-ranking further boosting performance~\cite{icralm:2023}. While re-ranking is preferred, employing any pre-trained retriever, particularly a text-based retriever, would be sub-optimal due to the inadequate information expected from ambiguous queries. Our work demonstrates the inadequacy of text-based retrievers for context retrieval and the necessity of more advanced retrieval models.

To address the lack of context inherent in under-specified queries, some studies have explored the use of CoT~\cite{cot:2022} mechanisms to generate text that closely approximates the semantic similarity of relevant context~\cite{qrrag:2023}. While CoT augmentation improves upon baseline methods, such as vanilla semantic search, CoT may potentially increase the input length to the LLM, which has a limited context window size. Additionally, studies have demonstrated that the placement of relevant information impacts LLM generation~\cite{lost-in-middle:2023}. Therefore, it is preferable to avoid increasing input sequence length if the same or better results can be achieved without query augmentation. Distillation-based query augmentation approaches have been proposed to address this problem~\cite{quill:2023}. Our work unveils that fine-tuning semantic search obviates the necessity for query augmentation while achieving comparable performance.

Recent studies have shown LLMs can act as zero-shot rankers through pairwise ranking prompting~\cite{llm-ranking:2023}. While addition of ranking for retrieval component has shown improvement in QA tasks, direct use of LLMs for the ranking task, in addition to plan generation, incurs twice the inference cost. We empirically show that our proposed lightweight context tuning method, LambdaMART~\cite{lambdamart:2010} based RRF~\cite{rrf:09}, outperforms both fine-tuning approach and GPT-4~\cite{gpt-4:23} based CoT Augmentation.

\section{Methodology}
\begin{figure*}[t!]
  \centering
  \includegraphics[width=\linewidth]{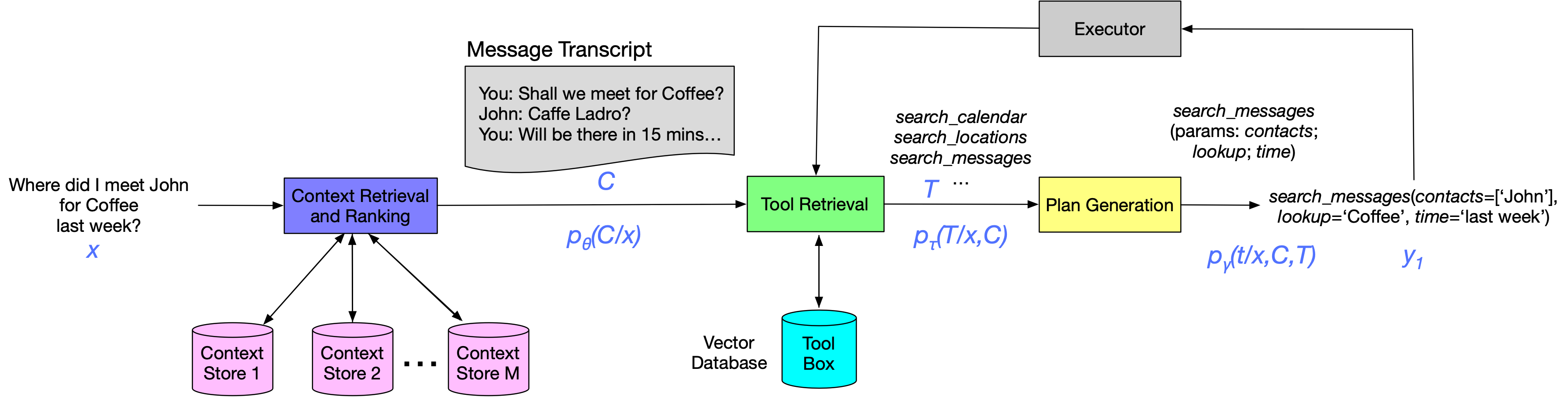}
  \caption{Context-tuned RAG pipeline illustrating end-to-end processing of a complex request with progressive plan generation.}
  \label{fig:ctrag}
\end{figure*}

Our experiments train and evaluate tool retrieval and planning with and without context tuning. Figure~\ref{fig:ctrag} illustrates how a context-seeking query uses context retrieval to enhance tool retrieval and plan generation.

\subsection{Data Generation}
Our study employed a data generation methodology using synthetic application data, aimed at simulating real-world scenarios for a digital assistant. The data encompasses 7 commonly used applications: mail, calendar, google, music, reminders, notes, and phone call. We generated this data using GPT-4, ensuring diversity in the dataset to reflect a wide range of user personalities. The synthetic dataset contained a diverse range of context items spanning various applications. A total of 791 distinct personas were synthesized, yielding 4,338 unique implicit queries for training and 936 implicit queries for evaluation.

Additionally, we developed a toolbox containing APIs for each of the applications we considered. This toolbox was created using in-context learning with GPT-4 and contained a total of 59 APIs distributed across the applications.

To simulate user interaction with a virtual assistant, GPT-4 was also utilized to generate realistic queries grounded in the application data. Following this, we employed GPT-4 to retrieve the appropriate tool from the generated toolbox in response to these queries. Finally, GPT-4 was used to resolve the tool's API with the correct parameters. This methodology provided a comprehensive and realistic dataset, essential for the evaluation of our context tuning approach in RAG-based planning systems.\footnote{Refer to Appendix~\ref{data_gen_appendix} for more details on data generation.}

\subsection{Context Tuning}
To compare various context retrieval methods, we employ both text-based and vector-based retrieval baselines. We simulate different context stores by structuring context data per persona and train models to perform federated search. We use query and persona meta-signals, such as frequency, usage history, and correlation with geo-temporal features, to perform retrieval. We evaluate context retrieval using the Recall@K and Normalized Discounted Cumulative Gain (NDCG@K) metrics.

\paragraph{\textbf{BM25}} For text-based search, we use an improved version of BM25, called BM25T~\cite{bm25:14}. 

\paragraph{\textbf{Semantic Search}} For vector-based search, we employ the widely adopted Semantic Search approach. We use GTR-T5-XL~\cite{gtr-t5-xl:21} to generate query and context item embeddings, which are then ranked using cosine similarity to select the top-K results. We evaluate both pre-trained and fine-tuned variants of this method.

\paragraph{\textbf{CoT Augmentation}} To enhance the likelihood of semantic alignment with pertinent contextual elements, we augment the under-specified or implicit query with GPT-4~\cite{gpt-4:23} generated CoT.\footnote{Please refer Appendix~\ref{cot_prompt_appendix} for the GPT-4 prompt used and Table~\ref{tab:cot_table} for CoT examples.} We evaluate both pre-trained and fine-tuned semantic search versions utilizing CoT.

\paragraph{\textbf{LambdaMART with RRF}} Reciprocal Rank Fusion (RRF)~\cite{rrf:09} is shown to outperform individual rank learning methods. To leverage this advantage, we propose a lightweight model that uses LambdaMART~\cite{lambdamart:2010} for initial ranking of data across context stores, followed by re-ranking using RRF.

\subsection{Tool Retrieval}
While advanced ranking models can enhance the recall of tool retrieval, we employ the pre-trained GTR-T5-XL model for semantic search using cosine similarity to retrieve the top-K tools. Extending the tool retrieval process to incorporate ranking should be a straightforward endeavor. We evaluate tool retrieval performance with and without context retrieval using Recall@K.

\subsection{Planner}
The planner's objective is to select the most appropriate tool from the retrieved tool list and generate a well-formed plan. A plan comprises an API call constructed using the chosen tool and parameters extracted from the query and retrieved context. We fine-tune OpenLLaMA-v2-7B~\cite{openllama-v2:2023} for plan generation. To assess the planner's performance, we employ the Abstract Syntax Tree (AST) matching strategy to compute plan accuracy. A hallucination is defined as a plan generated using an imaginary tool.

\section{Results}
\subsection{Context Retrieval}
\begin{table}[t]
\caption{A comparison of various Context Retrieval methods using Recall@K and NDCG@K metrics. 
The context-seeking query is used as input to perform a federated search across different context stores, after which semantic search or ranking is applied.}
\label{tab:cr_results}
\centering
\resizebox{\columnwidth}{!}{
\begin{tabular}{p{3cm}p{1cm}p{1cm}rp{1cm}p{1cm}r}\\
  \toprule
Retrieval Method & Recall@K & & & NDCG@K & & \\
 \cmidrule{2-7}
 & K=3 & K=5 & K=10 & K=3 & K=5 & K=10 \\
 \midrule
 BM25 & 11.35 & 13.47 & 14.92 & 56.45 & 52.33 & 50.91 \\
 Semantic Search & 23.74 & 25.38 & 26.99 & 65.44 & 64.31 & 64.02 \\
 CoT Augmentation & 71.77 & 85.61 & 94.41 & 93.67 & 91.78 & 88.40 \\
 Finetuned Semantic Search & 73.48 & 88.52 & 95.13 & 93.81 & 94.07 & 94.23 \\
 Finetuned w/ CoT Augmentation & 73.55 & 88.53 & 95.17 & 93.92 & 94.11 & 94.22 \\
 LambdaMART-RRF & \textbf{81.27} & \textbf{92.65} & \textbf{98.77} & \textbf{96.39} & \textbf{97.11} & \textbf{98.24} \\
\bottomrule
\end{tabular}}
\end{table}

Consistent with expectations, vector-based search surpasses text-based search, as shown in Table~\ref{tab:cr_results}. Nevertheless, both approaches struggle to retrieve relevant context for under-specified queries. Fine-tuned semantic search and CoT augmentation with pre-trained semantic search both significantly enhance retrieval performance. Notably, when fine-tuning is employed, CoT augmentation yields only marginal gains, suggesting that comparable improvements could be achieved without augmenting the input sequence with CoT.

Our proposed approach utilizing LambdaMART with RRF outperforms both fine-tuned semantic search and CoT augmentation. Additionally, we observe that for fine-tuned methods, both Recall@K and NDCG@K increase with K, whereas for pre-trained methods, NDCG@K decreases with an increase in K and Recall@K.

\subsection{Tool Retrieval}
\begin{figure}[t]
  \centering
  \includegraphics[width=\linewidth]{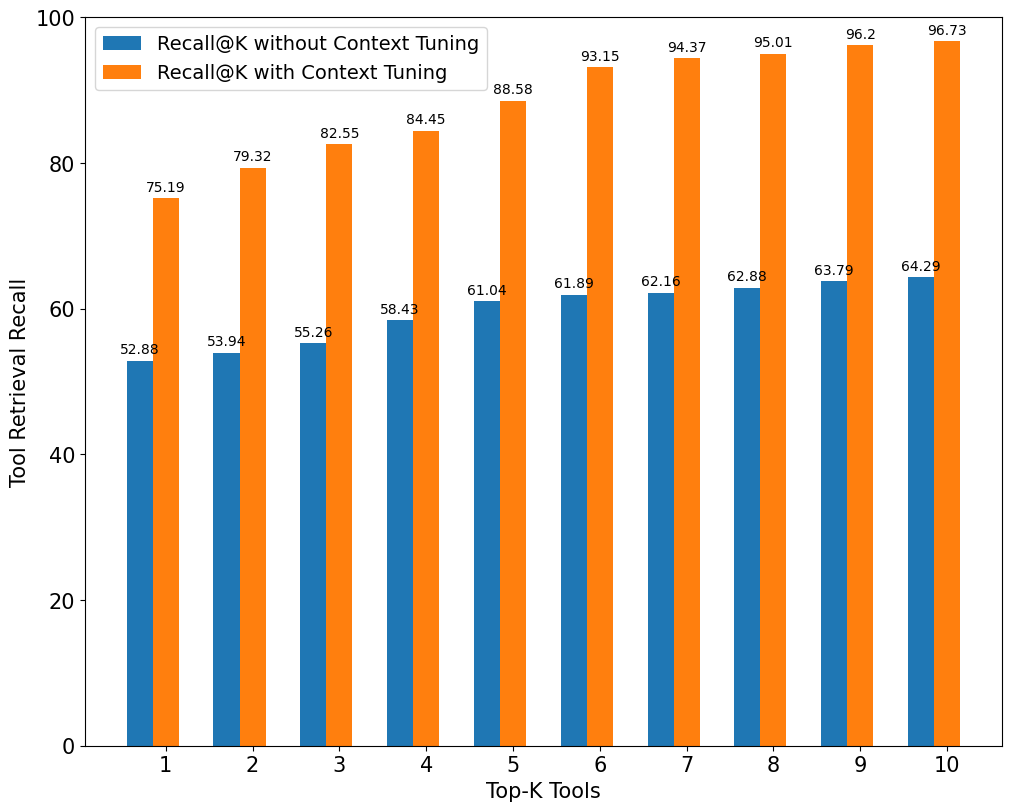}
  \caption{Evaluation of tool retrieval using Recall@k, with and without context tuning.}
  \label{fig:tr_recall@k}
\end{figure}
Figure~\ref{fig:tr_recall@k} illustrates the performance of tool retrieval using semantic search. Incorporating relevant context into tool retrieval consistently yields substantial gains across various K-values.

\subsection{Planner}
\begin{table}[t]
\caption{End-to-end planner evaluation both with and without context tuning. ``Lower Bound" excludes retrieval and performs direct plan generation while ``Upper Bound" assumes perfect context and tool retrieval.}
\label{tab:planner_results}
\centering
\resizebox{\columnwidth}{!}{
\begin{tabular}{p{2.5cm}p{2cm}cc}\\
  \toprule
Setting & AST-based Plan Acc $\uparrow$ & Exact Match $\uparrow$ & Hallucination $\downarrow$ \\
 \midrule
Lower Bound & 43.77 & 39.45 & 2.59 \\
\midrule
RAG-based Planner & 76.39 & 58.12 & 1.76 \\
Context-tuned RAG Planner & \textbf{85.24} & \textbf{67.33} & \textbf{0.93}\\
\midrule
Upper Bound & 91.47 & 72.65 & 0.85 \\
Context-tuned Upper Bound & 91.62 & 72.84 & \textbf{0.53}\\
\bottomrule
\end{tabular}
}
\end{table}
To establish the planner's lower bound, we remove the retrieval step, while the upper bound is set by directly utilizing context and/or tool labels, effectively employing oracle retrievers. Table~\ref{tab:planner_results} encapsulates the end-to-end evaluation of the fine-tuned planner, demonstrating that the context-tuned planner significantly outperforms the planner based on traditional RAG using semantic search. Notably, even when the correct tool is retrieved, incorporating relevant context in plan generation, as evidenced by the upper bound, helps in reducing hallucination.

\section{Conclusion}
Our work introduces context tuning, a novel component that enhances RAG-based planning by equipping it with essential context-seeking capabilities to address incomplete or under-specified queries. Through a systematic comparison of various retrieval methods applied to both lightweight models and LLMs, we demonstrate the effectiveness of context tuning in improving contextual understanding. Our empirical observations reveal that CoT augmentation enhances context retrieval when fine-tuning is not applied, while fine-tuning the retrieval model eliminates the need for CoT augmentation. Furthermore, we observe that context augmentation at the plan generation stage reduces hallucinations. Finally, we showcase the superiority of our proposed lightweight model using RRF with LambdaMART over the GPT-4-based system.

\section*{Limitations}
The current work does not utilize conversation history, which is crucial for handling explicit multi-turn instructions that contain anaphora or ellipsis. This limitation also hinders the model's ability to effectively process and respond to complex tasks that require multi-hop context retrieval. Additionally, the absence of conversation history impedes the model's ability to adapt to topic shifts that may occur throughout a dialogue.

Furthermore, the performance of the planner model is constrained by the length of the context window. While employing LLMs with longer context windows can enhance performance, it also increases model size and computational complexity. To address this limitation, incorporating context compression techniques could potentially improve end-to-end performance without incurring significant increases in model size.

Due to privacy constraints, we simulated real-world data by generating synthetic user profiles and personas that mirrored real-world use cases for a digital assistant.

\section*{Ethics Statement}
To safeguard privacy, this study exclusively utilizes synthetically generated data, eliminating the use of real user information under ethical considerations.

\section*{Acknowledgements}
We would like to thank Stephen Pulman, Barry Theobald and Joel Moniz for their valuable feedback.

\bibliography{custom}
\bibliographystyle{acl_natbib}

\appendix

\section{Data Generation Details}
\label{data_gen_appendix}

\subsection{Implicit Query Dataset}

For our experiments, we created a synthetic dataset to simulate realistic interactions across various applications commonly found with digital assistants. The dataset is structured to encompass a diverse range of contexts, representing different synthetic user activities and interactions.

\paragraph{Data Points:} A total of 791 unique personas were synthesized, covering seven key applications: Mail, Calendar, Google, Music, Reminders, Notes, and Phone Calls. The final dataset contained 4,338 train and 936 test data points.

\paragraph{Generation Method:} We utilized GPT-4 to generate the data. We ensured high diversity in the dataset is met through manual inspection, this is essential to accurately reflect a wide range of synthetic user personalities and interaction patterns.

\begin{figure}[ht]
    \centering
    \includegraphics[width=\linewidth]{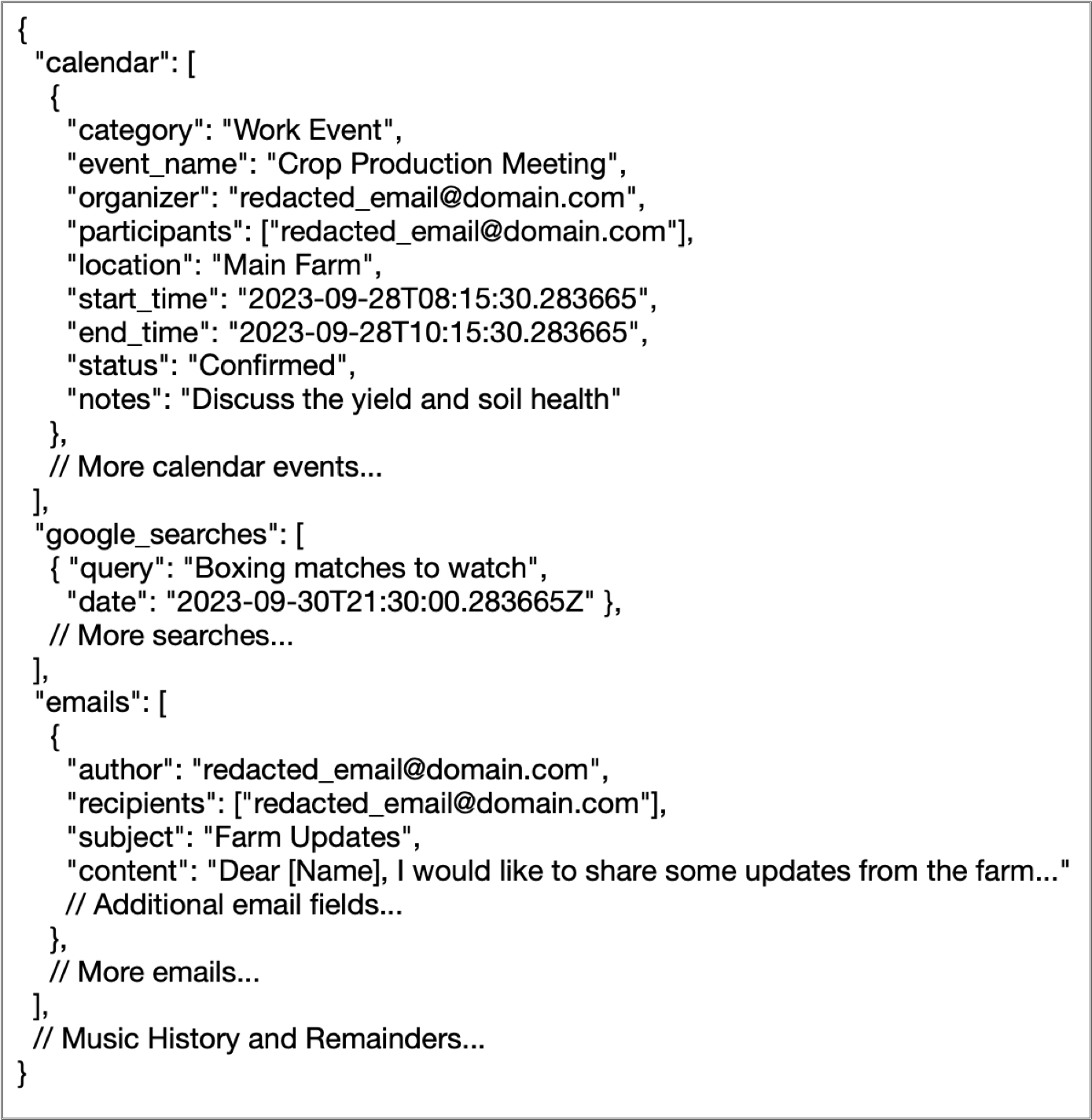}
    \caption{Snippet of a persona}
    \label{fig:json_data}
\end{figure}

\paragraph{Data Representation:} Each data point in the dataset contains multiple contextual information fields, relevant to the specific application and synthetic user's activity. An example of persona in JSON format is shown in Figure~\ref{fig:json_data}.

\label{subsec:context_distribution}

Table~\ref{table:context_distribution} shows the distribution of context items per application in our dataset.

\begin{table}[ht]
\centering
\begin{tabular}{|l|c|}
\hline
\textbf{Application} & \textbf{Avg. Context Items} \\
\hline
Mail & 2.93 \\
\hline
Calendar & 5.63 \\
\hline
Google & 9.57 \\
\hline
Notes & 2.23 \\
\hline
Music & 4.38 \\
\hline
Reminders & 4.81 \\
\hline
Phonecall & 2.34 \\
\hline
\end{tabular}
\caption{Distribution of context items per application.}
\label{table:context_distribution}
\end{table}

\subsection{Persona Data Creation Example Prompt}
\lstset{
  basicstyle=\small\ttfamily, 
  breaklines=true,            
  breakatwhitespace=false,    
}

\begin{lstlisting}
I'm working on generating synthetic data for a user (also known as persona) and the persona's
iPhone Data.

Here are the characteristics of the persona that we would like to generate the data for:
    
age: 22
favorite_music_genre: Pop
favorite_movie_genre: Romance
favorite_cuisine: Italian
favorite_sport: Tennis
profession: Software Developer
hobbies: ['Cooking', 'Swimming', 'Reading']

I want to generate data for ios App called Music with bundle id as com.apple.music.
Can you generate around 5 recently played songs

Instructions:
1. Today's date is 2023-12-07 11:18:19.028759, Please generate any times or dates in the past 15 days.
2. 'played_time' should be in yyyy-MM-dd HH:mm:ss.SSS format
        

Use the following schema:
The output should be formatted as a JSON instance that conforms to the JSON schema below.

As an example, for the schema {"properties": {"foo": {"title": "Foo", "description": "a list of strings", "type": "array", "items": {"type": "string"}}}, "required": ["foo"]}
the object {"foo": ["bar", "baz"]} is a well-formatted instance of the schema. The object {"properties": {"foo": ["bar", "baz"]}} is not well-formatted.

Here is the output schema:
```
{"$defs": {"MusicAppData": {"properties": {"recent_songs": {"items": {"$ref": "#/$defs/Song"}, "title": "Recent Songs", "type": "array"}, "current_playing": {"$ref": "#/$defs/Song"}}, "required": ["current_playing"], "title": "MusicAppData", "type": "object"}, "Song": {"properties": {"played_time": {"default": "", "title": "Played Time", "type": "string"}, "album_title": {"default": "", "title": "Album Title", "type": "string"}, "artist": {"default": "", "title": "Artist", "type": "string"}, "song_name": {"default": "", "title": "Song Name", "type": "string"}, "id": {"default": "", "title": "Id", "type": "string"}}, "title": "Song", "type": "object"}}, "properties": {"app_name": {"default": "", "title": "App Name", "type": "string"}, "app_bundle_id": {"default": "", "title": "App Bundle Id", "type": "string"}, "app_data": {"$ref": "#/$defs/MusicAppData"}}, "required": ["app_data"]}
```

Do not include any explanations, only provide a RFC8259 compliant JSON response following this format without deviation.

\end{lstlisting}

\subsection{Synthetic Toolbox Generation}
\lstset{
  basicstyle=\small\ttfamily, 
  breaklines=true,            
  breakatwhitespace=false,    
}
\begin{lstlisting}
You are an intelligent AI assistant tasked with generating APIs for iOS that can be used to interact with Applications. For example, if I ask you to generate APIs for Messages iOS Application, you would generate a comprehensive set of APIs that can perform any action on the app. Some examples below are:

api: read_message
description: Messages App's read_message API is used to read messages from a particular contact
arguments:
    - contact: contact from which the message was received

api: read_unread_messages
description: Messages App's read_unread_messages API is used to read all unread messages on your iPhone
arguments:
    -

api: send_message
description: Messages App's send_message API is used to send message to a particular contact
arguments:
    - text: text to be sent to the contact
    - contact: contact information

api: send_group_message
description: Messages App's send_group_message API is used to send a message to a list of contacts.
arguments:
    - text: text to be sent to the group
    - contacts: list of contacts in the group

api: search_messages
description: Messages App's search_messages API is used to search messages by text, recipient, sender.
arguments:
    - text: text to be searched.
    - recipient: search messages by recipient name
    - sender: Search messages by sender name

Similarly, can you generate the APIs for the following Application: {application}?
Do not include any explanations. Only provide the APIs in YAML format as above.
\end{lstlisting}
\label{subsec:toolbox_prompt_distribution}

The following table represents the distribution of APIs:

\begin{table}[ht]
\centering
\begin{tabular}{|l|c|}
\hline
\textbf{Application} & \textbf{APIs Count} \\
\hline
Music & 11 \\
\hline
Google & 10 \\
\hline
Notes & 9 \\
\hline
Mail & 8 \\
\hline
PhoneCall & 8 \\
\hline
Calendar & 7 \\
\hline
Reminders & 6 \\
\hline
\end{tabular}
\caption{Distribution of APIs generated by Synthetic Toolbox Generation}
\label{table:toolbox_distribution}
\end{table}

\subsection{Tool Retrieval}
\label{tr_prompt_appendix}

\lstset{
  basicstyle=\small\ttfamily, 
  breaklines=true,            
  breakatwhitespace=false,    
}

\begin{lstlisting}
I have the following toolbox defined with the available APIs: 
{tools}

For the following query: 
{query}

Suggest the most appropriate api? If there is no API available in the toolbox, then output default.
Only output the API name without any explanations
\end{lstlisting}

\subsection{Plan Resolution}
\label{pr_prompt_appendix}

\lstset{
  basicstyle=\small\ttfamily, 
  breaklines=true,            
  breakatwhitespace=false,    
}

\begin{lstlisting}
You are an intelligent AI Planner helping me come up with a plan and resolve the variables.

I have the following query:
{query}

I have selected the following tool to perform the task:
{tool}

Can you come up with fully resolved plan using the following schema?
{format_instructions}
\end{lstlisting}

\subsection{Prompt to generate CoT}
\label{cot_prompt_appendix}

\lstset{
  basicstyle=\footnotesize\ttfamily, 
  breaklines=true,            
  breakatwhitespace=false,    
  xleftmargin=0pt
}
\begin{lstlisting}
You are an expert in processing context-seeking or under-specified queries by finding missing context in the query. As an expert, your task is to generate concise chain of thought which when used to augment the context-seeking query, increases the semantic similarity of the updated query with relevant context items. Please only use the following context types: 'Mail', 'Calendar', 'Reminders', 'Notes', 'Photos', 'PhoneCall', 'Message', 'Messenger', 'Maps', 'Google Maps', 'Music', 'Spotify', 'Find My', 'Workout'; and do not create new context types.

Context-seeking Query: {query}

Your expert Chain of Thought:
\end{lstlisting}

Examples showing generated implicit queries along with CoT, context and plan labels are shown in Table~\ref{tab:cot_table}.

\begin{table*}
\caption{A sample of context-seeking or under-specified queries along with CoT produced by GPT-4. The columns for context and tools show labels for those retrieval tasks.}
\centering
 \label{tab:cot_table}
\scalebox{0.9}{
\begin{tabular}{l|l|l|l}
  \toprule
\textbf{Implicit Query} & \textbf{CoT} & \textbf{Relevant Context} & \textbf{Top-3 Relevant Tools} \\
 \toprule
\makecell[l]{When is my next \\ guitar lesson?} & \makecell[l]{Check the 'Calendar' for any \\ upcoming guitar lessons. \\ If not there, check 'Reminders' \\ for any alerts set about the lesson.} & \makecell[l]{The user has a reminder \\ titled ``Guitar Class"} & \makecell[l]{['Reminders', 'Calendar', \\ 'Notes']} \\
\midrule
\makecell[l]{I need to check my \\ diet plan again.} & \makecell[l]{I may have noted down the \\ diet plan in 'Notes'. If not \\ there, perhaps I saved a photo \\ of it in 'Photos'.} & \makecell[l]{The user has a note titled \\ ``Intermittent Fasting Plan." \\ The user also has an \\ image titled ``Keto Diet."} & \makecell[l]{['Photos', 'Notes', \\ 'Mail']} \\
\midrule
\makecell[l]{I'm running late.} & \makecell[l]{Check 'Calendar' for any \\ scheduled meetings. If so, verify \\ 'Maps' or 'Google Maps' to \\ gauge current traffic situation\\  and estimated time of arrival. \\ Use 'Messages' or 'Messenger' \\ or 'Mail' to inform the meeting \\ attendees that you are \\ ``running late".} & \makecell[l]{The user has an upcoming \\ meeting titled ``LLM \\ Discussion" organized by \\ ``John Doe."} & \makecell[l]{['Calendar', 'Mail', \\ 'Messages']} \\
\bottomrule
\end{tabular}}
\end{table*}

\end{document}